\documentclass{jetpl}
\usepackage{cite}
\twocolumn

\lat

\title{Dynamics of particles trapped by dissipative domain walls}

\rtitle{Dynamics of particles trapped by \ldots}

\sodtitle{Dynamics of particles trapped by dissipative domain walls}

\author{D.\,A.\,Dolinina\thanks{e-mail: d.dolinina@metalab.ifmo.ru}, A.\,S.\,Shalin, A.\,V.\,Yulin}

\rauthor{Dolinina D.A., Shalin A.S., Yulin A.V.}

\sodauthor{ Dolinina, Shalin, Yulin,}

\address{ITMO University,
197101 Saint-Petersburg, Russia}

\dates{}{}

\abstract{In this Letter we study the interactions of the dissipative domain walls with dielectric particles. It is shown that particles can be steadily trapped by the moving domain walls. The influence of the ratchet effect on particle trapping is considered. It is demonstrated, that the ratchet effect allows to obtain high accuracy in particle manipulation.}

\PACS{74.50.+r, 74.80.Fp}

\begin{document}

\maketitle

\section{INTRODUCTION}

Nonlinear localized structures have been attracting much attention in recent time because of the two reasons. The first one is fundamental interest to their rich variety in physical systems of different natures, including hydrodynamics, plasma physics, biology and nonlinear optics, see \cite{Water1, Plasma, Optics1, Cross}. And the second reason of high interest in nonlinear localized structures is their potential applications in many fields, including information optical processing \cite{Weiss2003,Kochetov}, optical fiber communications \cite{Mollenauer2},  and optical manipulation \cite{jetp,jetp2}.

One of the most interesting localized structures are domain walls connecting different stationary spatially homogeneous states. In general case a domain wall moves extending the area of one of the uniform states and thus after some time the expanding state fills the cavity. The direction and the velocity of the domain wall motion strongly depends on the pumping intensity. But there is a special value of pumping intensity characterized by zero velocity of the domain wall and it is called Maxwell point. Near the Maxwell point the domain walls are able to create different bound states, such as bright or dark solitons \cite{Yulin2008,Pesch2007}.

Another important effect of domain walls is reported in \cite{Yulin2016}. It is demonstrated that under biharmonical pumping the direction and the velocity of the domain wall can be controlled by changing only the mutual phase between the harmonics, it is so called <<ratchet effect>>.

In this Letter we suggest a new strategy of optical manipulation of small particles by dissipative domain walls. This problem is closealy related to the manipulation of the particles by dissipative bright solitons considered in \cite{jetp,jetp2}. This Letter is devoted to the formation, stability and the dynamics of the bound states of the particles and the domain walls. Special attantion is paid to the influence of the ratchet effect on the processes of particle capturing and on the possibility to use ratchet effect for nanoparticles manipulation.

\section{THE MATHEMATICAL MODEL}

We considered a nonlinear Fabry-Perot resonator pumped by the coherent light with a dielectric particle, located in the surface. Such resonators provide bistability and existence of bright solitons, see \cite{Szoke, Rosanov}. Also, it is known that the studied resonator, besides the bright solitons, is able to provide the existence of the dynamically stable domain walls, connecting two different spatially uniform states, see \cite{Yulin2016}. A particle on the surface of resonator is attracted in the area of higher intensity because of the gradient force \cite{Ashkin} and in \cite{jetp,jetp2} it is demonstrated that dissipative solitons in considered system are able to steadily capture particles and transport them in desirable direction.

The optical field of the considered resonator is described in the slow varying amplitude approach by the Schrödinger equation with the nonlinearity of saturable type, dissipation and pumping:

\begin{eqnarray}
    \dfrac{\partial}{\partial t} E - i C \dfrac{\partial^2}{\partial x^2} E + (\gamma + i \delta + i \dfrac{\alpha}{1 + |E|^2})E = \nonumber \\
    = (1 - f e^{-(x - \epsilon)^2/\omega^2})P,
    \label{eq:schrod}
\end{eqnarray}
where $C$ is diffraction coefficient, $E$ is a complex amplitude of optical field in the resonator, $P$ is an amplitude of laser pumping, $\gamma$ is decay rate, $\alpha$ is the nonlinearity coefficient; $\delta$ is laser detuning from resonant frequency, $\epsilon$ is coordinate of the nanoparticle. Parameter $\omega$ defines width of the shadow of a particle shadow situated at $x=\epsilon$, $f$ relates to the transparency of a particle: if $f = 0$, then the particle is transparent and if $f=1$ then the particle is opaque.
The motion of particle under the gradient force is described by the following equation for the particles' coordinate:
\begin{eqnarray}
    \dfrac{\partial}{\partial t} \epsilon = \eta \dfrac{\partial}{\partial x} |E(\epsilon)|^2,
    \label{eq:part}
\end{eqnarray}
In our model we use the typical assumption that the dragging force acting on the particle is proportional to the gradient of the intensity of the optical field, the coefficient $\eta$ accounts for the interaction strength. The motion of the particle is supposed to be viscous and thus it is described by first order ordinary differential equation. Let us note that for mathematical convenience we use the dimensionless variables.

The force  to the  field intensity gradient in the point of the particle location.

\begin{figure}[t]
\centering
\fbox{\includegraphics[width=\linewidth]{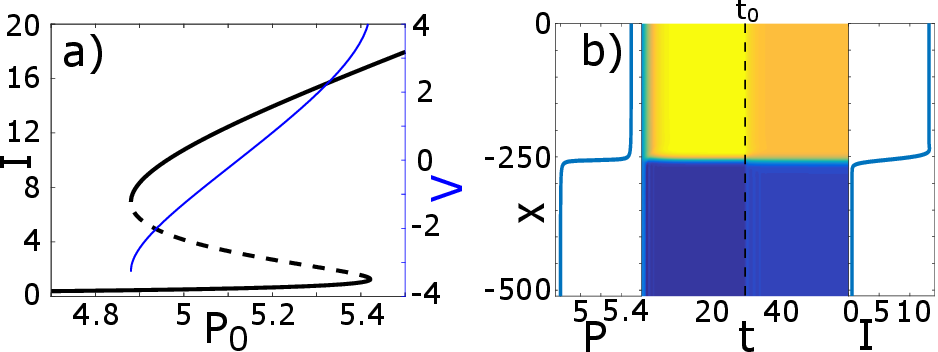}}
\caption{Fig.1 (a) The dependence of the intensity of spatially uniform states $I = |E|^2$ on pumping intensity $P_0$ is shown by black line and dashed part of the line shows dynamically unstable states (left vertical axes). The blue line shows the dependence of domain wall velocity on the pump intensity (right  vertical axes). The parameters are following: $\alpha = -10$, $\delta = -0.3$, $\gamma = 1$ and $C = 16$; (b) The formation of a domain wall from weak initial noise. First panel shows pumping field distribution at the beginning of numerical simulation till time $t_0$, at moment $t_0$ pumping field become uniform $P(x) = P_0 = 5.15$; second panel shows the field intensity distribution as a function of $t$; the last panel shows the formed domain wall.}
\label{fig0}
\end{figure}

The bifurcation diagram for spatially homogeneous stationary solutions under uniform pumping $P(x) = P_0$ is shown in fig.\ref{fig0}(a) by black line. The upper and the lower branches of the diagram are dynamically stable and the intermediate states are unstable. A domain wall in such system is a connection between the lower and the upper stable uniform states under the uniform pumping. The formation of the domain walls is the result of the complex interplay of the nonlinear Kerr effect, the dispersion, the driving force (the holding beam pumping the system), and the losses.

As it is mentioned before, in a general case the domain wall has non-zero propagation velocity . The velocity and the direction of the domain wall propagation can be controlled by the pumping intensity see  fig.\ref{fig0}(a) where the dependency of the domain wall velocity on the pump is shown by thin blue line (right axis). It should be noted, that there is a special value of pump intensity, when the domain wall is at rest. This specific pump is often referred as "Maxwell point".

We performed numerical simulations with the parameters insuring the existence of the domain walls under homogeneous pumping $P(x) = P_0$. In \cite{Yulin2016} it is shown that for the chosen set of parameters such nonlinear structures are always dynamically stable, regardless the pumping intensity. It was shown by direct numerical simulations that the domain walls can form from the weak initial noise if during the initial stage of the domain wall formation the pump intensity depends on coordinate, for example as it is shown in  fig.\ref{fig0}(b). Such type of pumping (first panel in fig.\ref{fig0}(b)) should provide the formation of the uniform state from the upper bifurcation branch in one part of the system and the state from the lowest bifurcation branch in the another part of the system. When the field intensity distribution becomes stationary the profile of the pump can be made spatially uniform without destruction of the domain wall, see the last panel in fig.\ref{fig0}(b). To proof that the domain walls observed in the direct numerical simulations are indeeed the connections of the spatially uniforms states we compared them against the stationary heteroclynic solutions found by Newton's method. This numerical results shows that the domain walls can be realized in the experiments. 

\begin{figure}[t]
\centering
\fbox{\includegraphics[width=\linewidth]{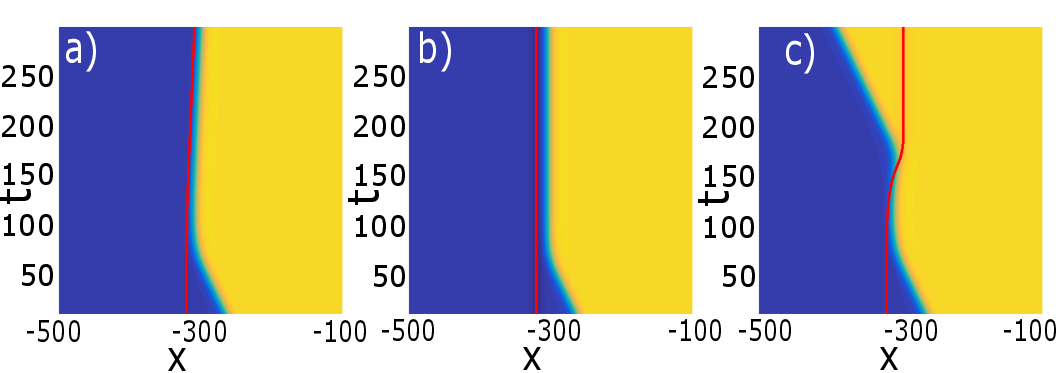}}
\caption{Fig.2 The interaction of the domain wall with a particle placed in the region of lower intensity under uniform time-independent pumping. Parameters are the same as in fig.\ref{fig0}, but $P_0 = 5.2$ and $\eta = 0.7$, $\omega = 10$ for all particles; (a) The domain wall changes direction of propagation because of the particle, $f = 0.025$; (b) The particle stops the domain wall, $f = 0.07$; (c) The particle is pulled in the region of stronger intensity, $f = 0.02$.}
\label{fig1}
\end{figure}

\section{THE INTERACTION OF DOMAIN WALLS WITH PARTICLES UNDER UNIFORM TIME-INDEPENDENT PUMPING}

In this section we study how the presence of particles affects the dynamics of domain walls discussed in the previous Section.
We focus on the dynamics of the domain walls with particle under uniform and time-independent pumping. Since the uniform states connected by the domain walls are not equivalent in the terms of intensities, the particle location relative to the wall is important. Let us begin with the case when the particle is placed in the area of lower field intensity. In this case several scenarios are possible.

If the decrease of the pump intensity caused by the shadow of the particle is sufficient, then the front changes the direction of propagation after collision with the particle. It can be explained by the following. The unperturbed pumping intensity is high enough and thus the system is switching from the lower spatially homogeneous stable state to the upper stable state by the motion of the domain wall connecting these states. The particle is placed far away from the domain wall and does not affect it. However, when the domain wall reaches the particle, the shadow created by the particle makes the pumping insufficient to maintain the upper state near the particle. In turn, the gradient force pulls the particle to the region of a stronger field intensity and the particle approaches the domain wall. Thus, the domain wall begins to slowly move in the opposite direction after the particle capture, as it can be seen in fig.\ref{fig1}(a).

It should be noted, that in case of the more opaque particle the domain wall can be almost stopped by the particle, see fig.\ref{fig1}(b). It happens because the pumping field near the particle is very weak  and the domain wall is not able to get close enough to the particle to pull it by the gradient force.

In case if the losses introduced by the particle in pumping intensity are not so strong, then the particle is pulled into the region of the strong field behind the front. After that the front again starts to move towards a homogeneous state of low intensity and the particle does not move, see fig.\ref{fig1}(c).

Now let us consider the case if the particle is placed in the area of higher intensity. In this case depending on the particle parameters also several scenarios are possible.

\begin{figure}[t]
\centering
\fbox{\includegraphics[width=\linewidth]{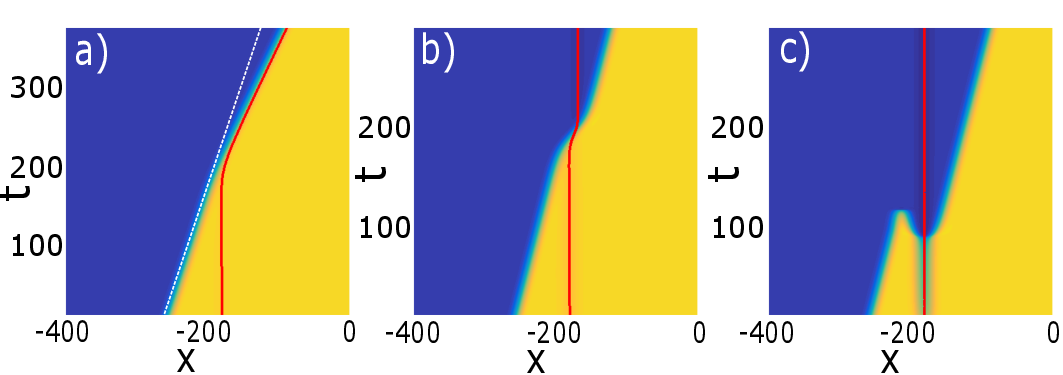}}
\caption{Fig.3 The interaction of the domain wall with a particle placed in the region of higher intensity under uniform time-independent pumping. Parameters are the same as in fig.\ref{fig1}, but $P_0 = 5.1$; (a) The particle is captured by the moving domain wall, $f = 0.005$. White dashed line shows the trajectory of the front without particle; (b) The particle breaks away from the domain wall, $f = 0.02$; (c) The particle creates two more domain walls, parameters are: $f = 0.07$ and $\omega = 10$.}
\label{fig2}
\end{figure}

If pump intensity decreasing is not significant then the capture of the particle by the moving dissipative domain wall is possible. After the capture of the particle  the velocity of the front increases because the particle's shadow leads to the decrease of the pump intensity. If the force dragging the particle is strong enough to maintain the velocity of particle equal to the increased velocity of the front, then steadily trapping of a particle is possible, see fig.\ref{fig2}(a).

However if the drop of the pump intensity caused by the presence of the particle is large enough then  the velocity of the domain wall increases beyond a threshold value and, as a result, the particle departs from the front. After that the domain wall restores it's initial speed without the particle, see fig.\ref{fig2}(b)

In case of even the more opaque particle placed in the region of the higher intensity the shadow created by the particle is able to switch the field from the upper state to the lowest in the vicinity of the particle. Thus, two more domain walls are created and the particle stays at rest in the region of lower field intensity, see fig.\ref{fig2}(c).

\section{THE INTERACTION OF DOMAIN WALLS WITH PARTICLES UNDER THE ACTION OF THE BIHARMONIC SIGNAL}

In this section we consider the influence of the biharmonic signal on the dynamics of the domain walls with particles interaction. The time-dependent spatially uniform pumping has the following form:
\begin{eqnarray}
P = P_0 + a_1 \sin(\Omega t) + a_2 \sin(2 \Omega t + \theta),
\label{eq:biharmonic}
\end{eqnarray}
where $P_0$ is time independent component of the signal, $\Omega$ is frequency of the first harmonic and $\theta$ is mutual phase difference between two harmonics.

\begin{figure}[t]
\centering
\fbox{\includegraphics[width=\linewidth]{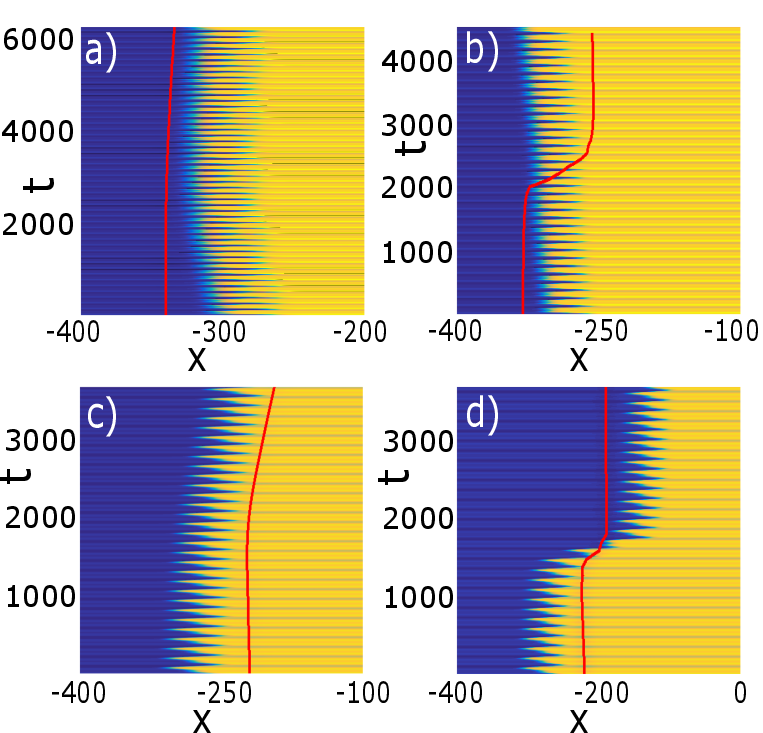}}
\caption{Fig.4 The interaction of the particle with the domain wall driven in motion by the ratchet effect. Parameters: $P = P_0 + a_1 \sin(\Omega t) + a_2 \sin(2 \Omega t + \theta)$, where $a_1 = a_2 = 0.1$ and $\Omega = 0.05$, other parameters are the same as in fig.\ref{fig1}; (a) Initially, the domain wall propagates in the direction of increasing of the higher state area because $\theta = 0$, but after collision with a particle ($f = 0.002$) changes the direction of propagation; (b) The particle is pulled in the area of higher intensity, parameters are the same as in (a), but $f = 0.0005$; (c) The domain wall propagates in the direction of increasing of the lower state area because $\theta = \dfrac{\pi}{2}$ and after collision with a particle ($f = 0.005$) increases the speed of propagation; (d) The particle breaks away from the wall and stays in the area of lower intensity, parameters are the same as in (a), but $f = 0.02$.}
\label{fig3}
\end{figure}

Under the action of biharmonic pumping signal it is possible to control the velocity of domain wall not only by changing the amplitude of pump but also by changing mutual phase of the harmonics, see \cite{Yulin2016}. This effect is especially important in the vicinity of the Maxwell point when the domain wall is at rest. If time-independent part of pumping intensity is close to Maxwell point, then by changing the mutual phase $\theta$ it is possible to change not only velocity of the domain wall, but also it's direction of propagation. In case if $\theta \approx 0$ the domain wall propagates in the direction of extension of the area of higher intensity, and in case if $\theta \approx \pi/2$ the domain wall moves in the opposite direction, see fig.\ref{fig3}.

Let us consider a case when the domain wall moves in the direction of increasing the lower spatially homogeneous state under the influence of the ratchet effect. When the domain wall reaches the particle and the particle is opaque enough then the wall changes the direction of propagation and the particle starts to move slowly with the domain wall, see \ref{fig2}(a). The effect is similar to the one shown in fig.\ref{fig1}(a)-(b).

In case if the particle is transparent enough the domain wall pulls the particle in the region of higher intensity and restores the initial average velocity, see fig.\ref{fig3}(b), what is similar to fig.\ref{fig1}(c).

Now let us consider the opposite case, when the domain wall moves in the direction of extension of the area of lower intensity because of the ratchet effect. From fig.\ref{fig3}(c) it is can be seen that after the domain wall reaches the particle and the particle is transparent enough then the domain wall captures the particle and they move together. Due to the viscous motion, the particle trajectory hasvery small oscillating component and it can be said that the particle is moving  with the velocity equal to the average velocity of the domain wall, see fig.\ref{fig3}(c).  The capture is similar to the one from fig.\ref{fig2}(a). If the transparency of the particle gets lower then the capturing regime is analogous to the case illustrated in  fig.\ref{fig2}(b). finally, if the particle is quite opaque, then particle breaks away from the wall.

Let us remark that we have studied the stability of the bound state of a partilce and a domain wall moving with a constant velocity. It was found that the eigenvalues describing the relaxation of the particle to its equilibrium point is pure real and this mean that the system does not have internal modes that can be resonantly excited because of the periodical variuation of the velocity or the profile of the domain wall. In other words no resonant excitation of the particle oscillations is possible in the most realistic case when the motion of the particle is viscous. It should be mentioned that if the inertia of the particle becomes of importance then the dynamics can become more complicated exhibiting different resonances. However these effects are out of the scope of the present Letter.

Let us emphasise the possible importance of ratchet effect for the manipulation of the nanopraticles by the nonlinear domain wall. The velocity caused by the ratchet effect can be made very low and  the particle shifts by a known distance over each of the period of the intensity oscillation. This way the shift of the particle can be controlled with high accuracy simply by counting of the number of the pump intensity oscillations.

\section{CONCLUSION}

Now let us briefly summarize the main results of the paper. The dynamics of a particle under the action of an optical nonlinear wave in a pumped dissipative resonator with saturable nonlinearity is considered. The considered nonlinear waves are the domain walls connecting two different dynamically stable spatially uniform states of the system. The dependence of the domain walls velocity on the pump is found numerically.

The interaction of the moving domain wall with the particle under the spatially uniform stationary pumping field is considered. It is shown that different scenarios of the mutual dynamics of the particle and the dissipative front are possible depending of the particle's transparency and its position in respect to the domain wall. It is demonstrated that the stable capture of the particle by the domain wall can be realized.

The influence of the ratchet effect on the domain wall with the particle interaction is considered. The possibility to use ratchet effect to set the particle into the controllable motion  is shown. It is worth mention that the ratchet effect can help to achieve high manipulation accuracy.

This work was supported by the Ministry of Science and Higher Education of Russian Federation (Goszadanie № 2019-1246). Also the work was partially supported by the Russian Foundation for Basic Research (Projects № 18-02-00414). The calculations of the fronts dynamics were supported by the Russian Science Foundation (Project № 18-72-10127).

\end{document}